**Author Correction: Superconductivity and normal-state transport in compressively strained $La_2PrNi_2O_7$ thin films**

Yidi Liu, Eun Kyo Ko, Yaoju Tarn, Lopa Bhatt, Jiarui Li, Vivek Thampy, Berit H. Goodge, David A. Muller, Srinivas Raghu, Yijun Yu and Harold Y. Hwang

In the version of this Article originally published, the Hall coefficient of $La_2PrNi_2O_7$ samples (P75, P78, P108) was reported with the wrong sign in **Fig. 4b** and **Extended Data Fig. 8**. This has now been corrected; the $La_3Ni_2O_7$ data from ref. 9 have been checked again and are unaffected.

Accordingly, the cuprate reference dataset in **Fig. 4b** has been updated to include Hall coefficient data from more heavily overdoped cuprates to compare with the corrected data. Ref. 25 has been changed from '25. Hwang, H. Y. *et al.* Scaling of the temperature dependent Hall effect in $La_{2-x}Sr_xCuO_4$. *Phys. Rev. Lett.* **72**, 2636–2639 (1994).' to '25. Tsukada, I. and Ono, S. Negative Hall coefficients of heavily overdoped $La_{2-x}Sr_xCuO_4$. *Phys. Rev. B* **74**, 134508 (2006).'

The corresponding text in the main manuscript has been revised from 'Furthermore, Hall resistivity measurements (**Extended Data. Fig. 6**) also indicate a similar temperature dependence and comparable values of Hall coefficient ($R_H(T)$) between $La_2PrNi_2O_7$ and overdoped $La_{1.75}Sr_{0.25}CuO_4$ (data from ref. [25], **Fig. 4b**).' to 'Furthermore, Hall resistivity measurements (**Extended Data. Fig. 6**) also indicate comparable values of negative Hall coefficient ($R_H(T)$) between optimized $La_{2-x}PrNi_2O_7$ and heavily overdoped $La_{1.6}Sr_{0.4}CuO_4$ (data from ref. [25], **Fig. 4b**). By contrast, the positive $R_H(T)$ in earlier $La_3Ni_2O_7$ samples[9] follows the behaviour of less overdoped $La_{1.75}Sr_{0.25}CuO_4$, perhaps suggesting a higher Ni oxidation state in our thin film $La_2PrNi_2O_7$.'.

These corrections do not alter our main conclusion in **Fig. 4** that "the normal-state transport behaviour resembles that of overdoped cuprates" but place $La_2PrNi_2O_7$ closer to the more heavily overdoped regime.

We thank Prof. Zhuoyu Chen for bringing this to our attention, and we deeply regret the error.

**List of Modifications**
**Main text**

- Starting from line 217:
  Before:
  Furthermore, Hall resistivity measurements (**Extended Data. Fig. 6**) also indicate a similar temperature dependence and comparable values of Hall coefficient ($R_H(T)$) between $La_2PrNi_2O_7$ and overdoped $La_{1.75}Sr_{0.25}CuO_4$ (data from ref. [25], **Fig. 4b**).

  After:
  Furthermore, Hall resistivity measurements (**Extended Data. Fig. 6**) also indicate comparable values of negative Hall coefficient ($R_H(T)$) between optimized $La_2PrNi_2O_7$ and heavily overdoped $La_{1.6}Sr_{0.4}CuO_4$ (data from ref. [25], **Fig. 4b**). By contrast, the positive $R_H(T)$ in earlier $La_3Ni_2O_7$ samples[9] follows the behaviour of less overdoped $La_{1.75}Sr_{0.25}CuO_4$, perhaps suggesting a higher Ni oxidation state in our thin film $La_2PrNi_2O_7$.

- Reference 25: We changed ref. 25 so that we can include Hall coefficient data of more overdoped cuprates to compare with our data.
  Before:
  25. Hwang, H. Y. *et al.* Scaling of the temperature dependent Hall effect in $La_{2-x}Sr_xCuO_4$. *Phys. Rev. Lett.* **72**, 2636–2639 (1994).

  After:
  25. Tsukada, I. and Ono, S. Negative Hall coefficients of heavily overdoped $La_{2-x}Sr_xCuO_4$. *Phys. Rev. B* **74**, 134508 (2006).

**Figures**

- **Fig. 4b**: Hall coefficient of bilayer nickelates.

<u>Before:</u>

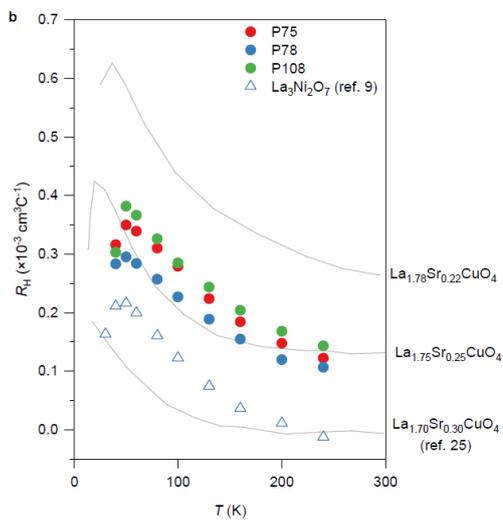

<u>After:</u>

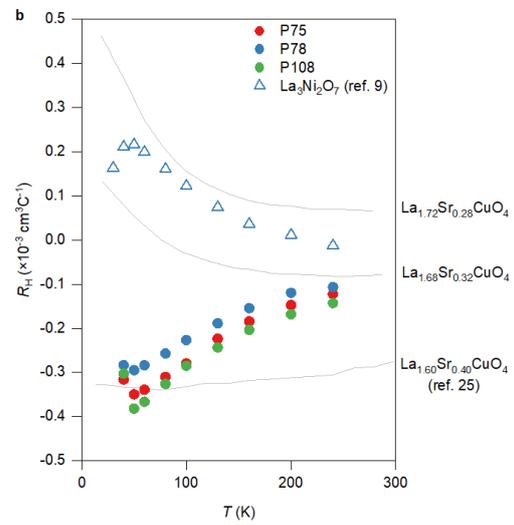

**Extended Data Display Items**

● **Extended Data Fig. 8**: Hall resistance vs. magnetic field for data in **Fig. 4b**. measurement

Before:

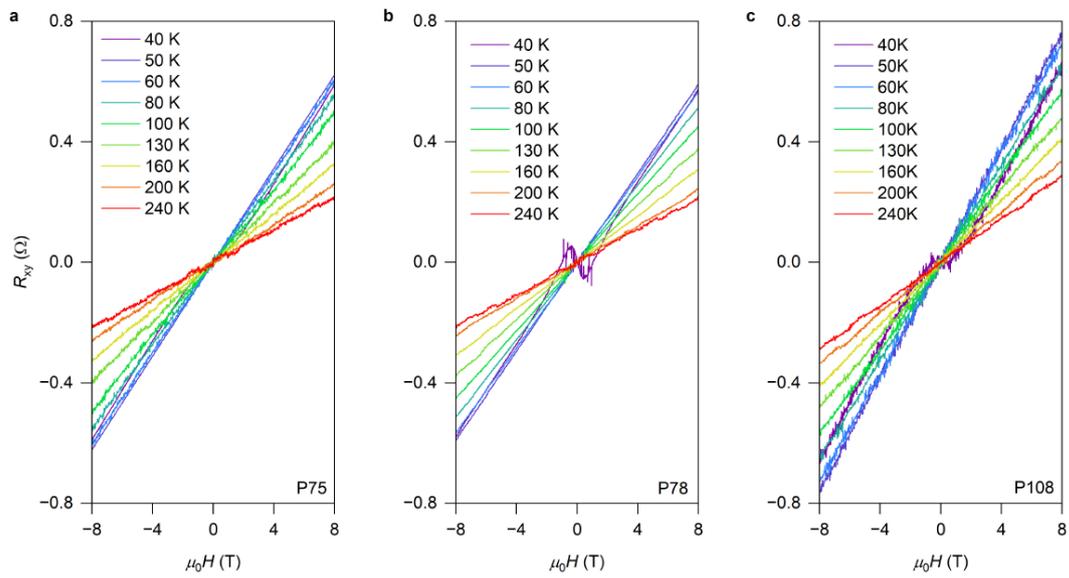

After:

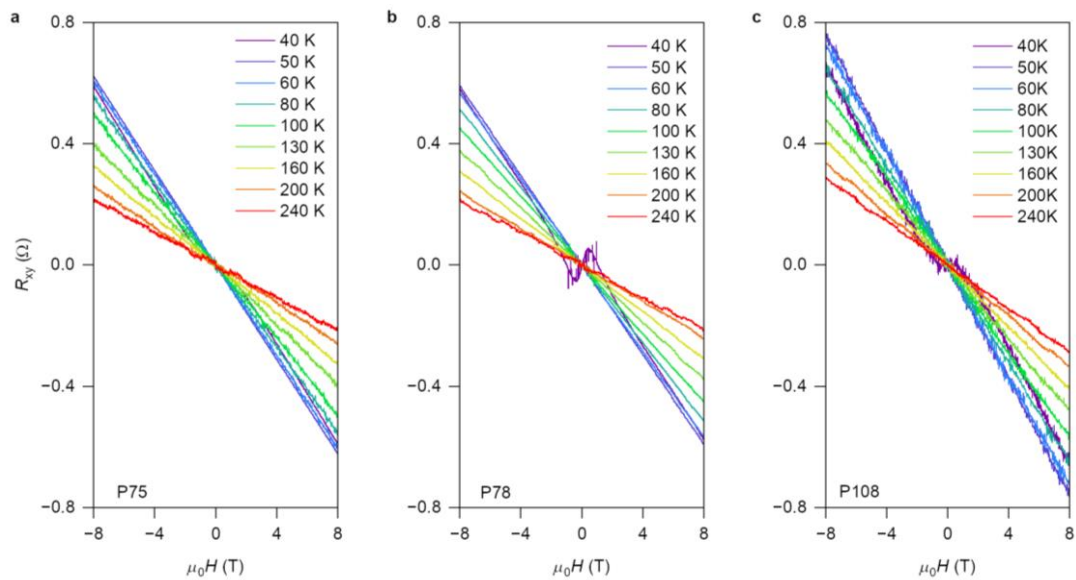

# Superconductivity and normal-state transport in compressively strained La$_2$PrNi$_2$O$_7$ thin films

Yidi Liu[1,2], Eun Kyo Ko[1,3], Yaoju Tarn[1,3], Lopa Bhatt[4], Jiarui Li[1,3], Vivek Thampy[5], Berit H. Goodge[6], David A. Muller[4,7], Srinivas Raghu[1,2], Yijun Yu[1,3]*, and Harold Y. Hwang[1,3]*

[1]*Stanford Institute for Materials and Energy Sciences, SLAC National Accelerator Laboratory, Menlo Park, CA 94025, USA*
[2]*Department of Physics, Stanford University, Stanford, CA 94305, USA*
[3]*Department of Applied Physics, Stanford University, Stanford, CA 94305, USA*
[4]*School of Applied and Engineering Physics, Cornell University, Ithaca, New York 14850, USA*
[5]*Stanford Synchrotron Radiation Lightsource, SLAC National Accelerator Laboratory, Menlo Park, CA 94025, USA*
[6]*Max Planck Institute for Chemical Physics of Solids, 01187 Dresden, Germany*
[7]*Kavli Institute at Cornell for Nanoscale Technology, Cornell University, Ithaca, New York 14850, USA*
*Email: yuyijun@stanford.edu, hyhwang@stanford.edu

**The discovery of superconductivity under high pressure in Ruddlesden-Popper phases of bulk nickelates has sparked great interest in stabilizing ambient pressure superconductivity in thin-film form using epitaxial strain. Recently, signs of superconductivity have been observed in compressively strained bilayer nickelate thin films with an onset temperature exceeding 40 K, albeit with broad and two-step-like transitions. Here, we report intrinsic superconductivity and normal-state transport properties in compressively strained La$_2$PrNi$_2$O$_7$ thin films, achieved through a combination of isovalent Pr substitution, growth optimization, and precision ozone annealing. The superconducting onset occurs above 48 K, with zero resistance reached above 30 K, and the critical current density at 1.4 K is 100-fold larger than previous reports. The normal-state resistivity exhibits quadratic temperature dependence indicative of Fermi liquid behaviour, and other phenomenological similarities to transport in overdoped cuprates suggest parallels in their emergent properties.**

**Main**

Ruddlesden-Popper (RP) phases of nickelates ($A_{n+1}Ni_nO_{3n+1}$, $A$ = La, Pr) are emerging as a new family of high-temperature superconductors[1–6], with bulk materials exhibiting superconductivity only under high pressure. In contrast, their thin-film counterparts, which can potentially benefit from the epitaxial strain imposed by the substrate[7,8], offer an additional degree of freedom to tune the lattice and electronic structure in ways that may stabilize superconductivity even at ambient pressure. Indeed, recent experiments on $La_{3-x}Pr_xNi_2O_7$ ($x$ = 0, 0.15) thin films, grown on SrLaAlO$_4$(001) [SLAO(001)] substrates with a nominal compressive strain of -2%, exhibit signatures of superconductivity with an onset temperature ($T_{c,onset}$) exceeding 40 K (ref. [9,10]). While highly promising, the broad, two-step-like superconducting transition, the low critical current density, and the presence of crystalline defects have impeded the study of the intrinsic superconducting and normal state properties.

In bulk bilayer nickelates, the complex stacking motifs of different RP phases[11–13] have been largely suppressed by $A$-site isovalent Pr substitution[5], and in thin films slight Pr substitution appears to similarly benefit superconductivity[10]. Here we have synthesized $La_2PrNi_2O_7$ on SLAO(001) by pulsed laser deposition (PLD, see **Methods**). Due to the very narrow growth window in the Ellingham diagram[9,14], we carefully optimized the growth conditions (temperature, oxygen partial pressure) to achieve the highest accessible film quality. A subsequent ozone annealing protocol, based on a bespoke system with resistivity feedback, was used to optimize the filling of oxygen vacancies while maintaining structural stability.

**Figure 1** summarizes the superconducting properties of our $La_2PrNi_2O_7$ films. As shown in **Fig. 1a**, we observe sharp superconducting transitions free of previously observed features of two-step transitions, with the maximum $T_{c,onset}$ exceeding 48 K (inset, with the definition of $T_{c,onset}$ explained graphically) and the highest zero-resistance temperature ($T_{c,0}$) surpassing 30 K. The improved superconducting transition is also reflected in the critical current density ($J_c$) measurements (**Fig. 1b**, $J_c$ defined in the caption). A 100-fold increase of $J_c$ (blue filled circles) compared to previous reported measures on bulk (red open triangles, calculated from ref. [2]) and $La_3Ni_2O_7$ films (green open circles, from ref. [9]) at 1.4 K provides strong evidence for the bulk nature of the observed superconductivity and is the highest value of $J_c$ reported across bulk and thin film bilayer nickelates. A clear diamagnetic response emerges around 13 K (**Extended Data Fig. 1**), similar to the temperature at which $J_c$ develops. Reciprocal space mapping (RSM) shows the sample is coherently strained to the substrate (**Extended Data Fig. 2a**). The decrease in in-plane lattice constant and the increase in $T_{c,onset}$ follow the same trend observed earlier (**Extended Data Fig. 2b**). Based on these observations, we suggest that the increased $T_{c,onset}$ may not be a direct consequence of Pr doping, which is further corroborated by the absence of $T_{c,onset}$ enhancement in bulk samples[5]. Instead, we attribute this to the improved sample quality achieved through Pr substitution, as well as optimized growth and ozone annealing.

Under magnetic field applied perpendicular (**Fig. 1c**) or parallel (**Fig. 1d**) to the film, superconductivity is gradually suppressed to lower temperatures without signatures of a second step in the transition. A plot of the upper critical fields ($H_{c,\perp}$ and $H_{c,\parallel}$) based on $T_{c,90\%}$ (defined as 90% of the resistance to the normal state near $T_{c,onset}$) is given in **Fig. 1e**, and the extracted zero-temperature Ginzburg-Landau in-plane coherence length is approximately 1.7 nm. The fitted superconducting thickness is around 4.9 nm, consistent with values obtained from Scherrer thickness analysis from X-ray diffraction (XRD) $\theta$-$2\theta$ symmetric scans (5 nm) and scanning transmission electron microscopy (STEM) measurements (5 nm, **Fig. 2e**). The observations presented here indicate the bulk nature of the superconductivity, and tend to rule out the possibility that the observed second step transition could arise from trilayer intergrowth domains, if any are present. In the following sections, we describe the key steps to achieve this.

**Growth optimization and structural characterisations**

The narrow growth window in the Ellingham diagram[9,14], the inherently complex bulk growth kinetics[11-13], and the additional strain energy accumulated during growth[9,15] collectively make the stabilization of highly crystalline bilayer nickelates particularly challenging. This difficulty is compounded by the limited ability to precisely control simultaneously both growth kinetics and thermodynamic conditions. However, we have successfully identified a stable growth window through fine-tuning of growth pressure and temperature. This progress likely benefits from a mechanism similar to the improved quality observed in bulk growth upon incorporating isovalent Pr substitution for La (ref. [5]).

**Figure 2a** presents a waterfall plot of XRD $\theta$-$2\theta$ symmetric scans of a series of samples grown under varying oxygen pressures ($p(O_2)$, at $\pm 1$ mTorr stability) with the substrate temperature at 660 °C. The bottom sample, grown at the highest $p(O_2)$, exhibits a minor (001) prepeak at ~23°, positioned to the left of the (006) peak of $La_2PrNi_2O_7$. This prepeak corresponds to the (La, Pr)NiO$_3$ phase. The top sample, grown at the lowest $p(O_2)$, reveals fringes centred around the substrate peak, indicating the predominance of the $(La,Pr)_2NiO_4$ phase. In between these two endpoints, the samples display a systematic evolution of XRD peaks, as evidenced by changes in their intensity (**Fig. 2b**) and position (**Fig. 2c**). The sample grown at $p(O_2)$ = 150 mTorr stands out with the highest intensity of (006) and (008) peaks (**Fig. 2b**), identifying the optimal coherent growth in this series. Additionally, within the growth series spanning 142 to 175 mTorr $p(O_2)$, the (006) peak positions remain relatively stable, indicating consistent structural characteristics across this range (**Fig. 2c**).

We then subjected all the samples to ozone annealing under the same conditions described in ref. [9], and subsequently examined the correlation between low-temperature transport and the XRD results (**Fig. 2d**, raw data shown in **Extended Data Fig. 3a**). We observed that all samples with the clustered (006) peak position have similar $T_{c,onset}$ while the trend of $T_{c,10\%}$ (following the definition of $T_{c,90\%}$) versus growth $p(O_2)$ closely followed the peak intensity evolution with $p(O_2)$ as shown in **Fig. 2b**. This indicates that the coherence of crystallinity in the sample is crucial in determining the fidelity of the superconductivity transition.

Direct visualization of the atomic structure in the optimized samples grown at $p(O_2)$ = 150 mTorr using STEM (**Fig. 2e**) confirms that the crystallinity is indeed improved compared to earlier $La_3Ni_2O_7$ films[9]. However, higher order RP phases are still occasionally visible. This may explain why the onset temperatures for mutual inductance and critical current measurements remain lower than $T_{c,onset}$. Additionally, $J_c$ is still orders of magnitude lower than that of infinite-layer nickelates[16] and cuprates[17]. Another interesting observation is that there is frequently one layer of $(La,Pr)_2NiO_4$ at the interface before $La_2PrNi_2O_7$ grows, a phenomenon not seen in ref. [10]. We suspect that this occurrence is sensitive to different substrate treatment/reconstructions and growth methods, and the electrostatic boundary conditions imposed. However, this contrasting phenomenon suggests that the superconductivity reported in thin films is largely unaffected by the interfacial electronic structure, but rather arises from the mechanical boundary conditions imposed by epitaxial strain.

While octahedral tiling in bulk $La_2PrNi_2O_7$ leads to an orthorhombic structure, films grown on SLAO reinforce a tetragonal symmetry due to substrate-induced constraints (**Extended Data Fig.2a**). In the bulk phase, the onset of superconductivity under pressure coincides with an orthorhombic-to-tetragonal structural transition[5,18,19]. This raises the question of whether the symmetry plays a role in inducing superconductivity in bilayer nickelates. To date, two different atomic structures have been proposed for the superconducting phase under pressure[1,18], differing in both space group assignments and the oxygen positions in the $NiO_2$ plane. This highlights the need for a more systematic structural study of superconducting nickelates, with thin films that can superconduct under ambient pressure offering valuable opportunities for this investigation.

**Post-growth ozone annealing with *in situ* transport feedback**
A curious fact is that for both infinite-layer nickelate and bilayer nickelate thin films, as-grown samples do not exhibit superconductivity. Superconductivity only emerges after a post-growth soft chemistry treatment, either involving the removal[16] or refilling[9,10,20] of oxygen. This two-step procedure inherently adds complexity to the stabilization of superconductivity.

To overcome this challenge, we have constructed an ozone annealing setup with *in situ* transport capabilities and tunable ozone concentration ($w(O_3)$) (see **Extended Data Fig. 4**). For all samples annealed in this setup, we traced the resistivity change over the annealing time ($t_{anneal}$) and concluded the annealing when the resistivity saturated (inset of **Extended Data Fig. 5a** as an example). By determining $t_{anneal}$ in this way, we minimized the uncertainty caused by potential crystallinity fluctuations between different samples and ensured the best oxygen stoichiometry achievable in the experiments (**Methods** and **Extended Data Fig. 6**). Varying $w(O_3)$ and annealing temperature ($T_{anneal}$), we constructed a phase diagram in **Fig. 3a**, which is reminiscent of the Ellingham diagram referenced during the growth stage[9,14]. We found that the combination of high $w(O_3)$ and low $T_{anneal}$ drives the as-grown bilayer films towards the higher-order RP phases (**Fig. 3a**, red filled dots; see also ref.[9]), while the combination of low $w(O_3)$ and high $T_{anneal}$ can keep the bilayer structure intact (**Fig. 3a**, blue open and filled dots). Thus, a phase boundary (dashed line in **Fig. 3a**) can be sketched following the Clausius-Clapeyron form [$w(O_3) \propto \exp(k/(T_{anneal})$, $k$ is a constant]. Among all the samples retaining the bilayer structure, we used filled blue dots to mark those exhibiting superconductivity in the subsequent temperature dependent resistivity ($\rho(T)$) measurements. We observed that below the phase boundary, the superconducting samples are clustered in upper right region (high $w(O_3)$ and high $T_{anneal}$), while the non-superconducting ones are located in the lower left region (low $w(O_3)$ and low $T_{anneal}$). This observation is consistent with the notion that sufficient kinetic activation is necessary to achieve complete oxygenation.

It is worth noting that $w(O_3)$ here only reflects the ozone concentration at the outlet of the ozone generator at room temperature, while the actual ozone concentration at the sample surface is difficult to quantify due to the steep temperature dependence of the ozone lifetime $\tau$ (ref.[21]). To provide qualitative indication of the oxidizing environment, we show a colour-shaded relative oxidizing power, which is proportional to $w(O_3) \cdot \tau$, in the background of **Fig. 3a** as a reference.

With all of the above information at hand, we then proceeded to optimize ozone annealing conditions by varying $w(O_3)$ and $T_{anneal}$, and then examining the $\rho(T)$ curves. The samples we used here are typically not fully optimized; however, we found that samples with more defects are also more sensitive to changes in annealing conditions, which actually benefitted our study. **Figure 3b** and **c** show the $T_{c,onset}$ versus $w(O_3)$ and $T_{anneal}$, respectively, with the other factor kept constant (see **Extended Data Fig. 3b** and **c** for original $\rho(T)$ curves). We observe that both $w(O_3)$ and $T_{anneal}$ have optimal values, with the optimal combination of $[T_{anneal}, w(O_3)]$ (labelled with a green × in **Fig. 3a**) located in the upper right corner of the annealing condition (labelled with an orange + in our previous study[9].

The key challenge of the post-annealing process lies in optimising competing factors—introducing more oxygen without triggering a transformation into the locally thermodynamically stable phase. Our final optimized ozone annealing pathway is plotted as the green trajectory in **Fig. 3a**. An example run on Sample P75 demonstrates the time-dependent profiles of $\rho$, $w(O_3)$, and $T_{anneal}$ in **Extended Data Fig. 5a-c**, respectively. The rationale behind this pathway is that: 1) it crosses the phase boundary quickly, between 150 and 320 °C , taking 3 minutes for warming up and 7 minutes for cooling down as depicted in **Extended Data Fig. 5c** (compared to 12 minutes and 55 minutes, respectively, in our previous study); 2) the change in $w(O_3)$ during warming up and cooling down (**Extended Data Fig. 5b**) ensures it crosses the phase boundary at the lowest possible temperatures, where kinetics are slower; 3) ozone annealing is performed in a single step, as we observed that superconductivity consistently degrades after a second annealing, with some samples even transforming into the perovskite phase after multiple rounds of stepwise annealing due to excessive time spent near the phase boundary.

A final technical question concerns the degradation of samples after growth or ozone annealing and how to preserve superconducting samples. Our systematic study reveals a subtle yet irreversible degradation for as-grown samples over a timescale of months (**Extended Data Fig. 7a-b**), along with rapid oxygen loss for ozone-annealed samples within days under ambient conditions (**Extended Data Fig. 7c**). To address this, we developed a cryogenic storage method, as detailed in the **Methods** section and **Extended Data Fig. 7d-f**.

**Normal state transport**
As shown in Fig. 1, our combined optimization strategy has led to the observation of high-quality superconducting transitions in transport measurements. The normal state transport is also much improved both based on the low resistivity (down to 50 μΩ·cm at 60 K, Fig. 1a) and higher residual resistivity ratio (RRR, defined as $\rho(300\ K)/\rho(60\ K)$, which exceeds 5) compared to earlier reports[9,10].

We note that the small $\rho$ in our La$_2$PrNi$_2$O$_7$ films is comparable to that of overdoped cuprates, and the parallel-resistor formula (PRF)[22,23]

$$\frac{1}{\rho(T)} = \frac{1}{\rho_0 + AT^2} + \frac{1}{\rho_{MIR}}$$

provides an accurate fit to our data (**Fig. 4a**), whereas neither $\rho \sim T + T^2$ or $\rho \sim T^n$ (with $1 < n < 2$)[24] yields a satisfactory fit. Here, $\rho_0 + AT^2$ represents the typical resistivity of a Fermi liquid (with $\rho_0$ the residual resistivity and the term $A$ characterizing electron-electron scattering). The term $\rho_{MIR}$ corresponds to the resistivity at the Mott-Ioffe-Regel limit. The fitted $\rho_{MIR}$ values are 0.7 and 1.7 mΩ·cm for the two samples, reasonably well matched with the value fitted in cuprates[23]. Our analysis suggests that the normal state of the film exhibits Fermi liquid behaviour, similar to that observed in overdoped cuprates. Furthermore, Hall resistivity measurements (**Extended Data. Fig. 6**) also indicate comparable values of negative Hall coefficient ($R_H(T)$) between optimized La$_2$PrNi$_2$O$_7$ and heavily overdoped La$_{1.6}$Sr$_{0.4}$CuO$_4$ (data from ref. [25], **Fig. 4b**). By contrast, the positive $R_H(T)$ in earlier La$_3$Ni$_2$O$_7$ sample[9] follows the behaviour of less overdoped La$_{1.75}$Sr$_{0.25}$CuO$_4$, perhaps suggesting a higher Ni oxidation state in our thin film La$_2$PrNi$_2$O$_7$.

Related studies on bulk bilayer nickelates under pressure have indicated a linear $T$-dependent $\rho(T)$ in samples with the highest $T_c$ under optimal pressure[2], along with a contribution of $T^2$-like dependence at lower pressure just above $T_c$ (ref. [2]). While systematic bulk $R_H$ measurements remain unavailable, existing data[26] suggest a pressure-dependent decrease of $R_H$ measured at 90 K, with values comparable to our data. Further studies, including both high-pressure bulk experiments and thin films under varied epitaxial strain, would be useful to determine the relationship between bulk hydrostatic pressure and biaxial compressive strain in terms of the impact on electronic structure.

We close by discussing the implications of the normal state transport in the context of the cuprate superconductors, and address the extent to which our measurements suggest a similarity between the cuprates and the bilayer nickelate films. Quantum chemistry considerations suggest the possibility of substantial differences between the two systems, since Ni electrons are in a $d^{7.5}$ configuration unlike the cuprates, which start with $d^9$ electrons in a partially filled $d_{x^2-y^2}$ orbital. Therefore, one cannot *a priori* rule out the importance of a partially occupied $d_{z^2}$ band in this system, and many theoretical studies support its leading role[27-37]. While density functional studies predict that the $d_{z^2}$-dominant $\gamma$ band is further removed from the Fermi surface with compressive strain, resulting in a more cuprate-like electronic structure[38,39], $d_{z^2}$ electrons may play a more crucial role in a strong-coupling limit[33,40-44], where the interplay of strong onsite repulsion and Hund's coupling may result in the formation of a Mott insulator of $d_{z^2}$ electrons[41,45]. In this case, their interaction with holes in the $d_{x^2-y^2}$ band would suggest a behaviour more akin to heavy fermion materials. Furthermore, related nickelates with Ni electrons in the $d^8$ configuration exhibit a high spin state in which the $d_{z^2}$ spins play an essential role[46]. If such closely related materials inform the present bilayer nickelate, they would indicate a substantial difference between the cuprate and bilayer nickelate superconductors. Within this context, the similarity of $\rho(T)$ and $R_H(T)$ observed here with overdoped cuprates is striking and leads us to speculate that the normal state transport—and perhaps the correlations responsible for superconductivity—may be dominated by the $\alpha$ and $\beta$ bands, primarily originating from Ni $d_{x^2-y^2}$ orbitals[47-50]. In this framework, as estimated in ref. [48], the $\alpha$ and $\beta$ bands shift from a heavily overdoped regime to a more 'optimal' doping level under pressure. Whether a similar response occurs under biaxial compressive strain presents an interesting avenue for exploration. Finally, we caveat that the effects of doping in the cuprates and that of compressive strain in the bilayer nickelates are likely not identical, which makes a precise comparison with the cuprates difficult. Nevertheless, our observations do suggest an underlying similarity in the emergent behaviour of both the cuprate and nickelate superconductors.

**Acknowledgments**

We thank Y. Deng, K. Lee, Y. Lee, Y. Lv, J. May-Mann, F. Theuss, B. Y. Wang, Y.-M. Wu, and J. J. Yu for discussions and assistance. Y.L., E.K.K., Y.T., J.L., S.R., Y.Y., and H.Y.H. acknowledge support from the U. S. Department of Energy, Office of Basic Energy Sciences, Division of Materials Sciences and Engineering (Contract No. DE-AC02-76SF00515), as well as SuperC and the Kavli Foundation (aspects of magnetic characterization). Work at the Stanford Nano Shared Facilities (SNSF) RRID:SCR_023230 is supported by the National Science Foundation under grant ECCS-1542152. L.B. and D.A.M. acknowledge support from National Science Foundation (DMR-1719875), NSF PARADIM (DMR-2039380) and the Weill Institute and the Kavli Institute at Cornell University. X-ray measurements were carried out at the SSRL, SLAC National Accelerator Laboratory, supported by the US Department of Energy, Office of Science, Office of Basic Energy Sciences (contract no. DE-AC02-76SF00515). B.H.G. acknowledges support from the Max Planck Society and Schmidt Science Fellows in partnership with the Rhodes Trust.

**Contributions**

Y.Y. and H.Y.H. conceived and designed the project. Y.L. and E.K.K. synthesised the films. Y.Y. built the ozone annealing setup and studied ozone annealing effects with Y.L. and Y.T. Y.L. characterized and studied the transport properties of films with the assistance of Y.T. L.B., B.H.G. and D.A.M. measured and analysed STEM images. J.L., Y.L., Y.T., and V.T. performed the RSM measurements. Y.Y., Y.L., S.R., and H.Y.H. analysed data and wrote the manuscript with input from all authors.

**Competing financial interests**

No author has any competing financial interests.

**Figure captions**

**Figure 1 | Superconductivity in thin film La$_2$PrNi$_2$O$_7$. a**, $\rho(T)$ curves of La$_2$PrNi$_2$O$_7$ films (samples P75, P78, P81, and P87) grown on SLAO(001) substrates. The inset zooms in on the region near $T_{c,0}$. **b**, $J_c(T)$ for sample P75, with bulk La$_3$Ni$_2$O$_7$ data from ref. [2] and thin film La$_3$Ni$_2$O$_7$ data from ref. [9] included for comparison. The inset shows a log-log plot of electric field ($E$) vs. current density ($J$) measured at 1.4 K, 3-16 K at 1 K intervals, and 18 K (right to left). $J_c$ is defined as the value of $J$ when $E$ drop to the noise floor. **c-d**, $\rho(T)$ under various magnetic fields applied perpendicular (**c**) and parallel (**d**) to the thin film measured on sample P75. **e**, Upper critical fields ($H_{c,\perp}^{90\%}$ and $H_{c,\parallel}^{90\%}$) extracted using $T_{c,90\%}$ values are represented by solid circles, with solid lines indicating Ginzburg-Landau fits[51].

**Figure 2 | Growth optimization and structural characterisation. a**, XRD $\theta$-$2\theta$ symmetric scans of La$_2$PrNi$_2$O$_7$ films grown under different $p(O_2)$ at 660 °C. SLAO(001) substrate peaks are marked by ♦. **b**, Intensity of (008) and (00$\underline{12}$) peaks of La$_2$PrNi$_2$O$_7$ films as functions of $p(O_2)$. Vertical error bars represent instrumental noise. **c**, Position of the (00$\underline{12}$) peak for La$_2$PrNi$_2$O$_7$ films plotted against $p(O_2)$. **d**, $T_{c,\text{onset}}$ and $T_{c,10\%}$ of La$_2$PrNi$_2$O$_7$ films as functions of $p(O_2)$. **e**, ADF-STEM image of an optimal La$_2$PrNi$_2$O$_7$ sample on SLAO(001) along the [110] zone axis of the substrate. Across most of the field of view, a layer of (La,Pr)$_2$NiO$_4$ is grown on the surface of the substrate prior to the growth of La$_2$PrNi$_2$O$_7$. Atomic models are overlaid on the left and right side of the image. Near the centre, also overlaid with an atomic model, a small region reveals the formation

of a bilayer structure between the substrate and film before the growth of $La_2PrNi_2O_7$. The blue and red shaded octahedra represent bilayer and monolayer structures, respectively. Atom species are colour-coded as indicated in the legend.

**Figure 3 | Ozone annealing optimisation. a,** Experimentally obtained phase diagram of $T_{anneal}$ vs. $w(O_3)$ for ozone annealing. Red filled dots represent films transformed to higher-order RP phases, while blue dots correspond to films remaining in the bilayer phase. Filled (open) blue dots are films with (without) superconducting transitions. The black dashed line represents the estimated boundary between the bilayer phase and the perovskite phase. The orange "+" denotes the annealing condition from ref. [9], while the green trajectory and "×" denote the optimized annealing protocol from this study. **b-c,** $T_{c, onset}$ vs. $T_{anneal}$ and $w(O_3)$, respectively, while holding the other parameter constant.

**Figure 4 | Normal state transport of $La_2PrNi_2O_7$ films. a,** $\rho(T)$ of two thin film $La_2PrNi_2O_7$ samples (P75 and P81). Dashed lines are fit with the PRF between 55 to 280 K as described in the main text. **b,** $R_H(T)$ for $La_2PrNi_2O_7$ samples and $La_3Ni_2O_7$ (ref. [9]) compared with overdoped $La_{2-x}Sr_xCuO_4$ (replotted from ref. [25]).

## Methods

### Sample preparation and XRD characterization

Polycrystalline La$_2$PrNi$_2$O$_7$ and single crystalline SrTiO$_3$ were used as targets during the deposition process. The polycrystalline La$_2$PrNi$_2$O$_7$ target was synthesized from La$_2$O$_3$, Pr$_6$O$_{11}$ and NiO powder, sintered 2 times: 1000 °C for 12 hours, 1100 °C for 48 hours. The La$_2$PrNi$_2$O$_7$ (~ 5 nm) and $SrTiO_3$ films (1 unit cell) were deposited by PLD with a KrF excimer laser (wavelength 248 nm). During the $La_2PrNi_2O_7$ and $SrTiO_3$ film growth, the substrate temperature was maintained at 660 °C and 620 °C, respectively, under oxygen partial pressure $p(O_2)$ of 150 mTorr. The pulsed laser repetition rate during the deposition was 5 Hz for $La_2PrNi_2O_7$, and 3 Hz for $SrTiO_3$. The laser fluence was 0.56 $J/cm^2$ with a spot size of 3.3 $mm^2$. The $SrLaAlO_4(001)$ substrates (MTI Corporation) were sonicated in acetone and IPA for 1 minute each. After growth, XRD data were measured using a monochromated Cu $K_{a1}$ source ($\lambda = 1.5406$ Å). RSM measurements were performed at beamline BL17-2 of the Stanford Synchrotron Radiation Lightsource, using a photon energy of 8.33 keV at room temperature. Diffraction signals were collected using a PILATUS3 100K-M detector.

**Transport and mutual inductance measurements**

40 nm thick Au electrodes were deposited using electron beam evaporation and then electrodes were bonded to ceramic chip carriers with aluminium wires by an ultrasonic wire bonder before performing ozone annealing. During ozone annealing, electrical transport was measured inside a tube furnace (**Extended Data Fig. 3**). Subsequent low-temperature magnetotransport measurements were performed using $^4$He cryostats. 2-coil mutual inductance measurements were performed in the same setup as in ref. [52].

**Optimal timing of ozone annealing**

Experimentally, resistivity during ozone annealing serves as a key indicator for determining whether the oxygen vacancies have been sufficiently filled, even though the exact density of vacancies remains unknown. This approach is based on the assumption that a stoichiometric sample should reach a resistivity minimum, as minimal scattering from oxygen vacancies would be present. Indeed, in multiple samples, we observed that when resistivity reaches saturation before terminating the ozone annealing, the superconducting transition of this sample is sharp and complete. In contrast, if annealing is terminated before resistivity saturation, the superconducting transition remains broad and incomplete (**Extended Data Fig. 6**). This observation is self-consistent with the above assumption ,and based on this, resistivity saturation is established as our standard for optimal ozone annealing.

**Sample degradation study and storage methods**

a.  Phase stability of as-grown thin films (Stability of the bilayer phase in storage conditions)

For as-grown samples without post-growth ozone annealing, XRD $\theta$-$2\theta$ symmetric scans (**Extended Data Fig. 7a**) show no noticeable signs of degradation after five months of storage in ambient conditions compared to freshly grown samples. However, we observed degradation in samples that underwent ozone annealing after being stored for more than two months, compared to those annealed immediately after growth, as shown in **Extended Data Fig. 7b**. Specifically, two specimens of sample P75 underwent identical ozone annealing conditions but with a 5-month interval. The specimen annealed after five months exhibited a significantly broadened superconducting transition, a trend that has been consistently observed across multiple samples. This observation suggests that the bilayer thin films undergo slow but irreversible degradation over time. Given the ultrathin nature of the films, and the formal thermodynamic instability of the 327 phase in ambient conditions, this level of degradation is not unexpected.

b.  Stability of ozone-annealed thin films (Stability of oxygen content in storage conditions)

We have consistently observed that the resistivity of samples increases immediately after ozone annealing and eventually reaches an insulating state within a few weeks (**Extended Data Fig. 7c**, and **e**), consistent with previous reports (refs. 9 and 10). This resistivity trend can be largely reversed by performing another round of ozone annealing, indicating that the primary factor is a reversible change in oxygen stoichiometry. However, repeated ozone annealing gradually degrades the overall sample quality, as noted in the main text. The temperature below which oxygen stoichiometry remains stable—reflected in resistivity measurements—is around 200 K, similar to the value reported in ref. 10.

c.  Methods to prevent degradation in bilayer nickelate samples

To mitigate degradation in both bilayer structure and oxygen content, we explored potential strategies to stabilize resistivity, including storage under high oxygen pressure, and at low temperature. A summary is provided in **Extended Data Fig. 7e**. We found that storing samples at liquid nitrogen temperature fully stabilized their resistivity, maintaining a constant value for at least a month (**Extended Data Fig. 7d**). A practical and convenient storage method (inspired by ref. [53]) is illustrated in **Extended Data Fig. 7f**.

**STEM measurements**

To prepare cross-sectional lamellas, we used a standard focused ion beam (FIB) lift-out procedure using a Thermo Fisher Helios G4 UX FIB.  We performed STEM imaging in ADF mode using a Cs-corrected Thermo Fisher Scientific (TFS) Spectra 300 X-CFEG 300 kV with a probe convergence semi-angle of 30 mrad.  Inner/outer collection angles in ADF geometry were 77/200.

**Data availability**

The source data presented in the figures are provided with this paper. Any additional data that support the findings of this study are available from the corresponding author upon reasonable request.

**Methods-only references**

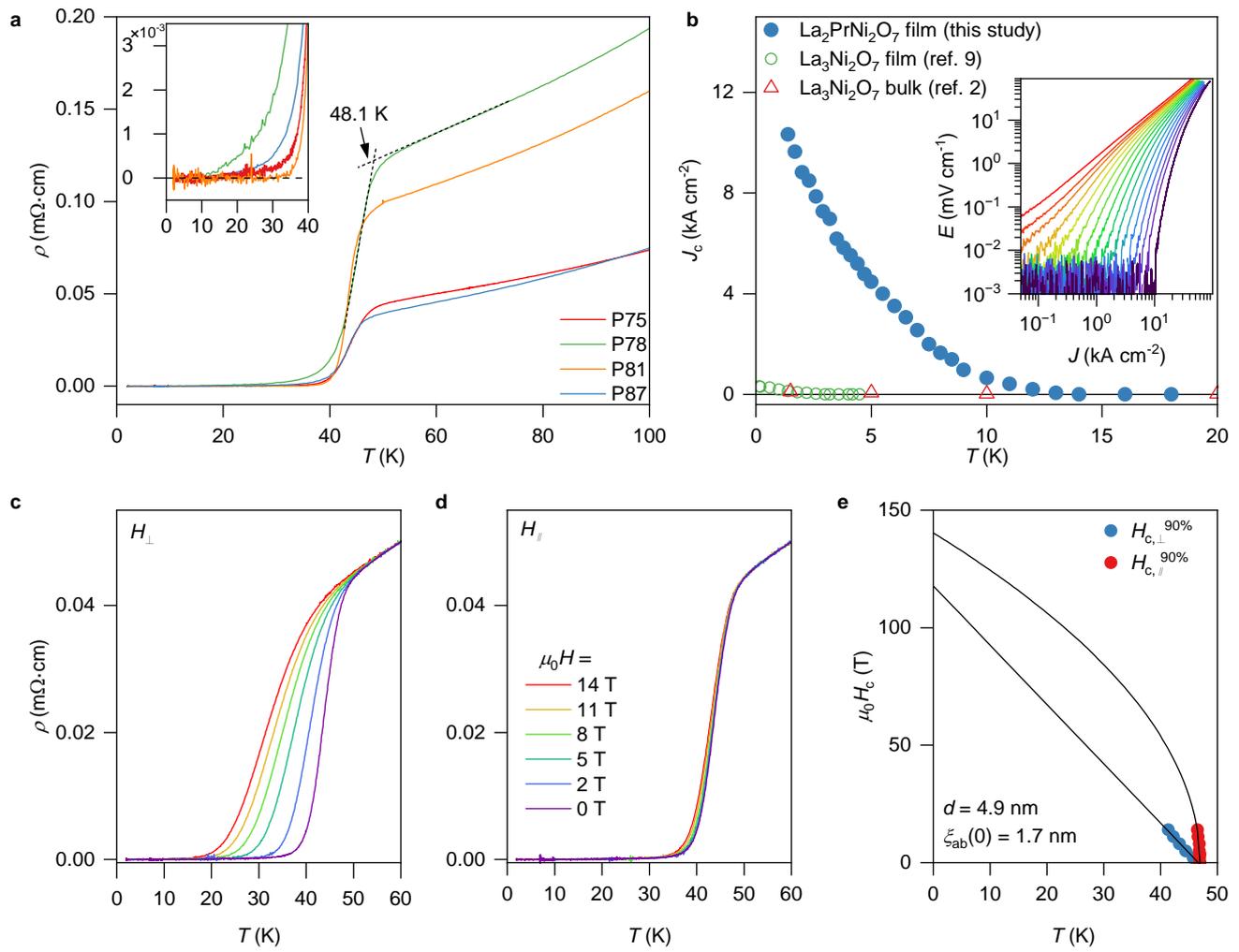

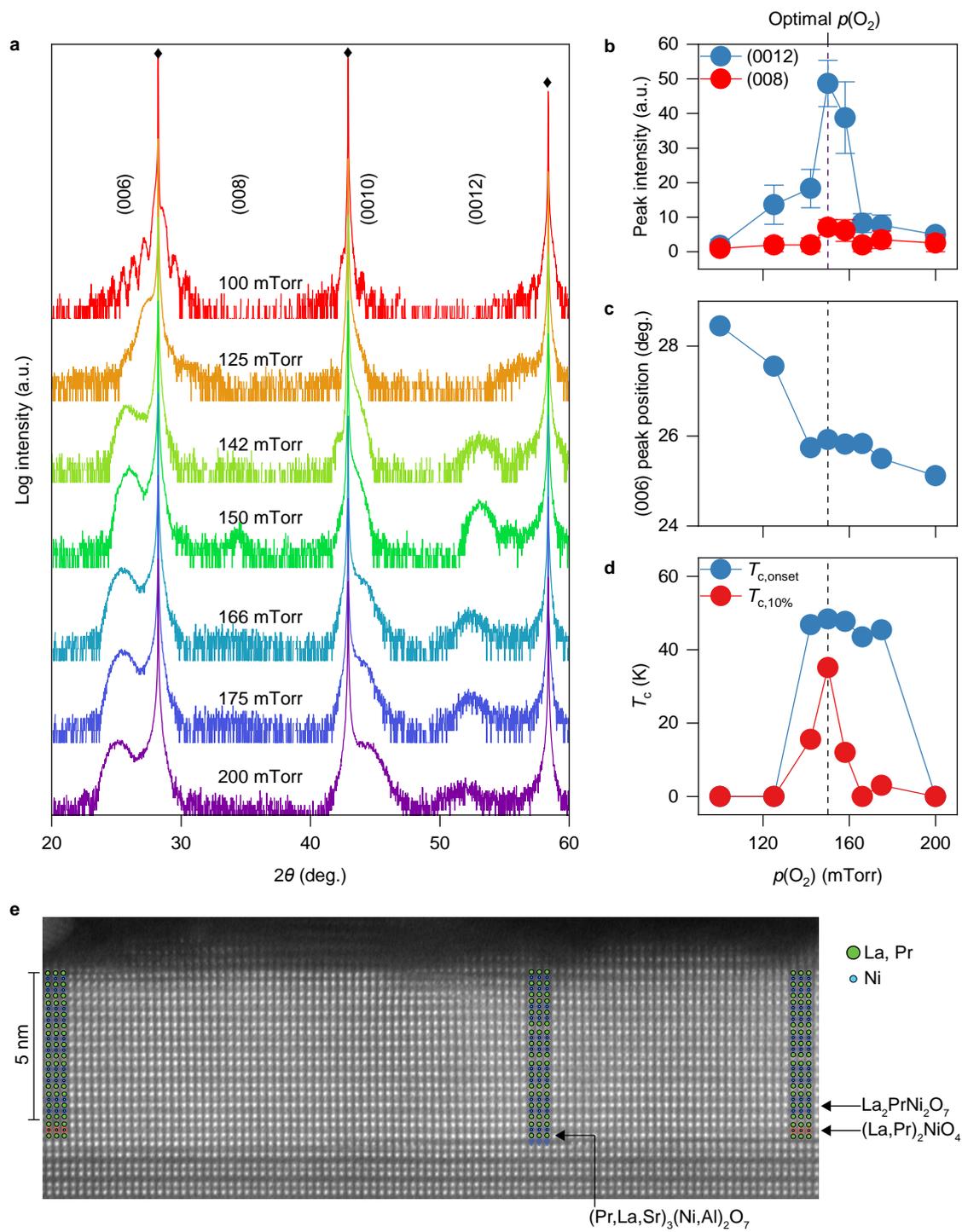

b **Optimal** $p(O_2)$

c

d

e

- **La, Pr**
- **Ni**

5 nm

La₂PrNi₂O₇
(La,Pr)₂NiO₄

(Pr,La,Sr)₃(Ni,Al)₂O₇

Liu et al., Figure 2

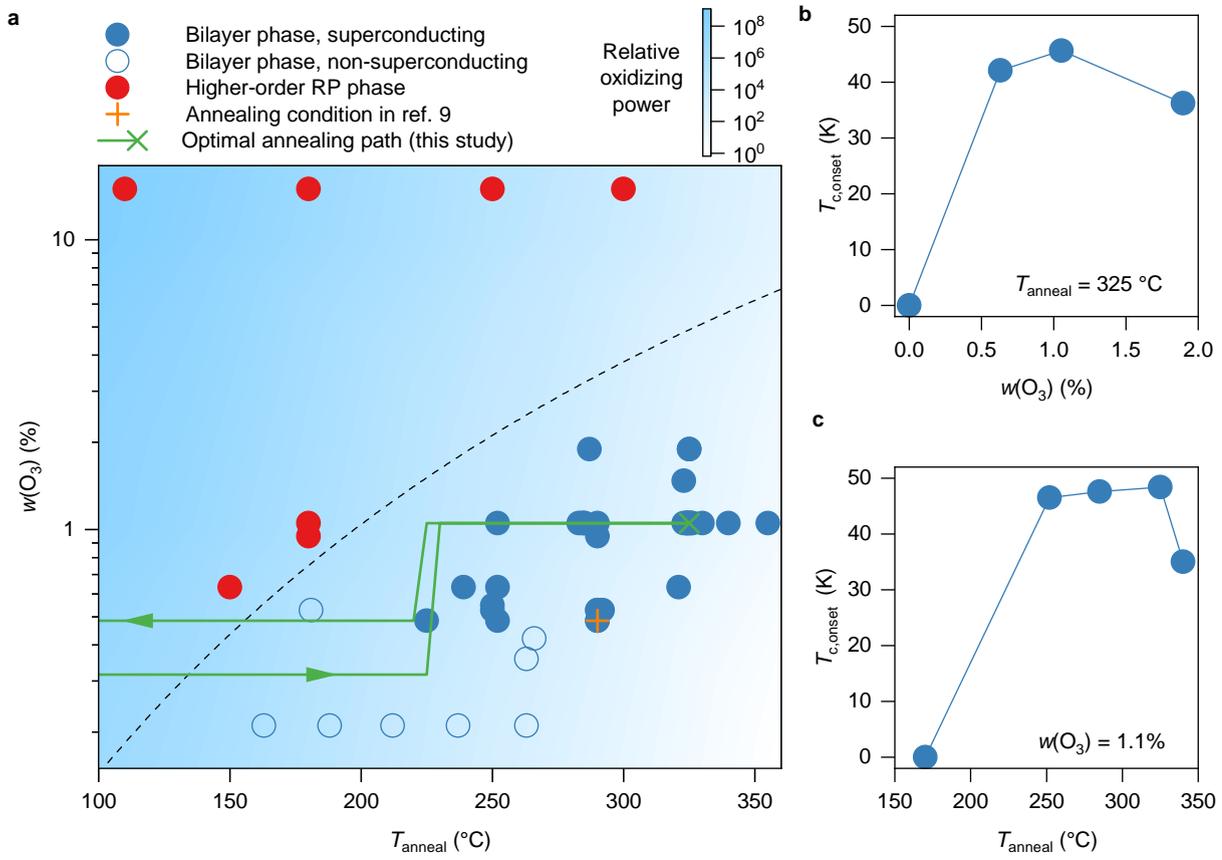

Liu et al., Figure 3

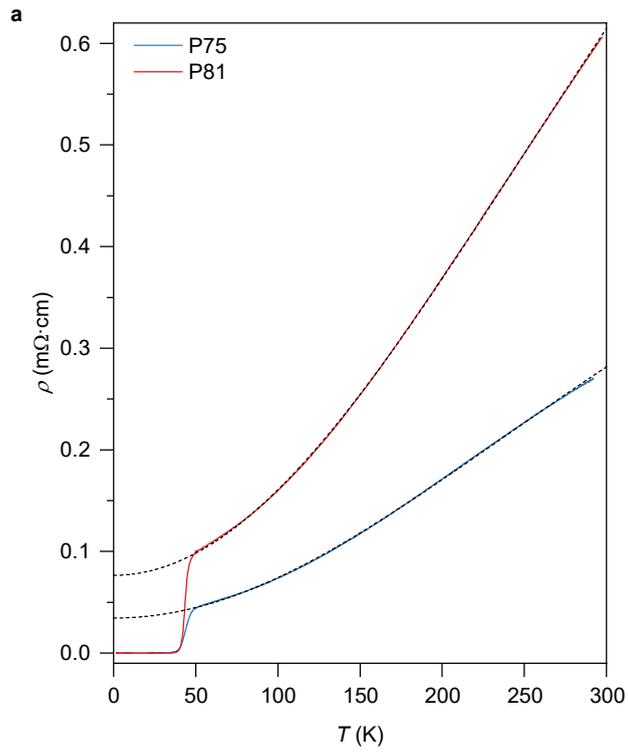

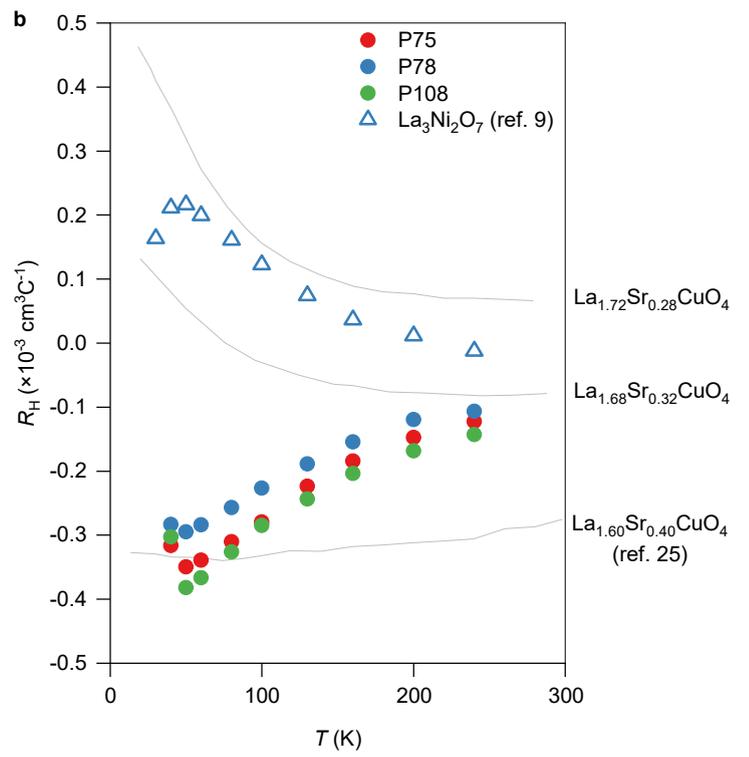

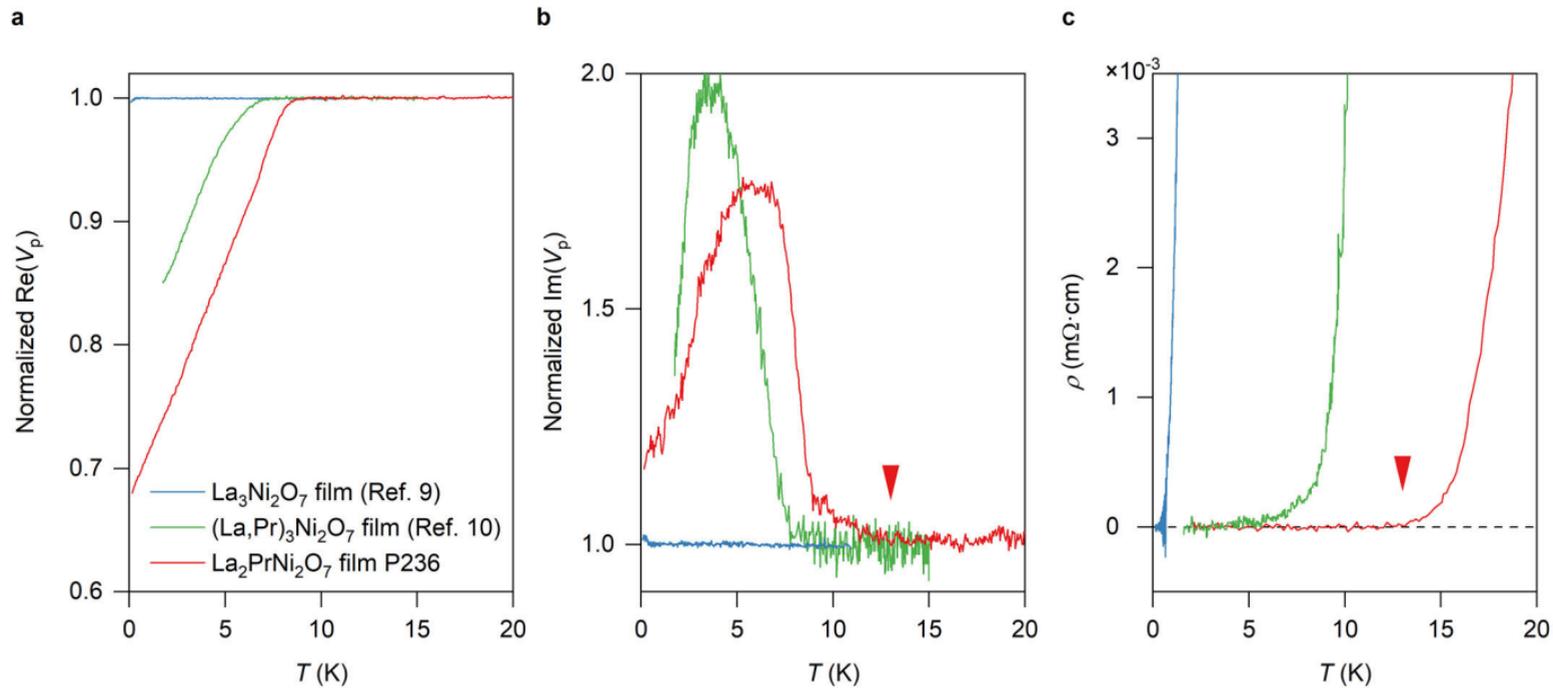

Liu et al., Extended Data Figure 1

**a**

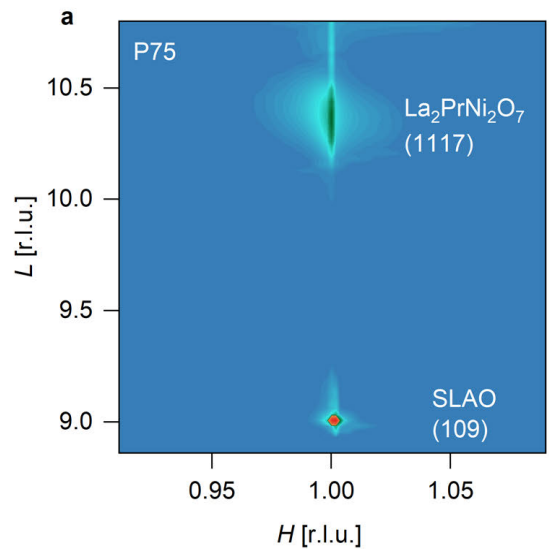

**b**

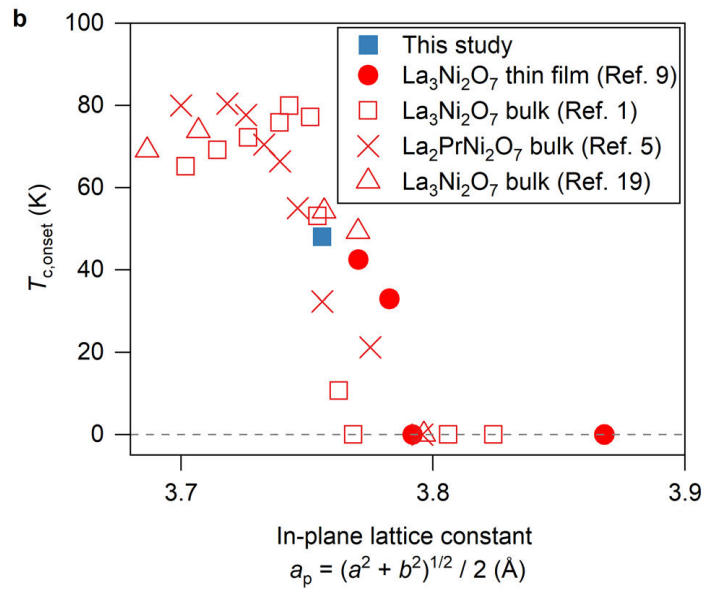

Liu et al., Extended Data Figure 2

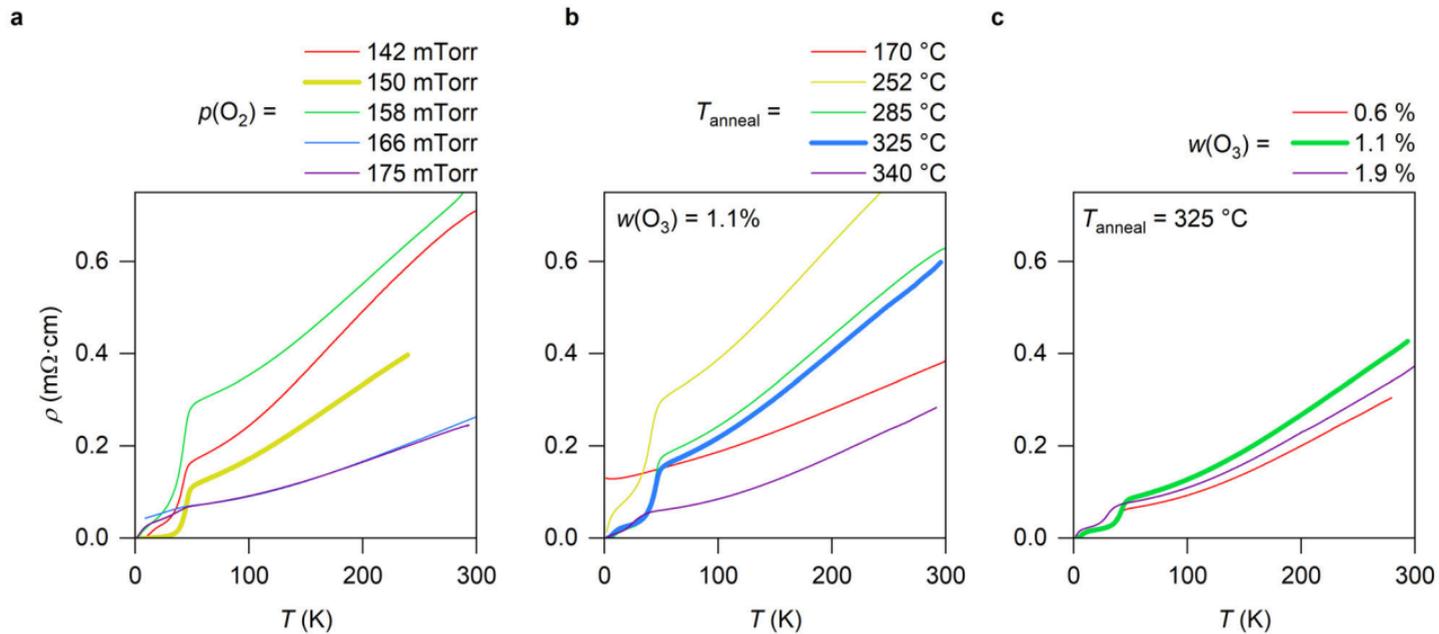

Liu et al., Extended Data Figure 3

**a**

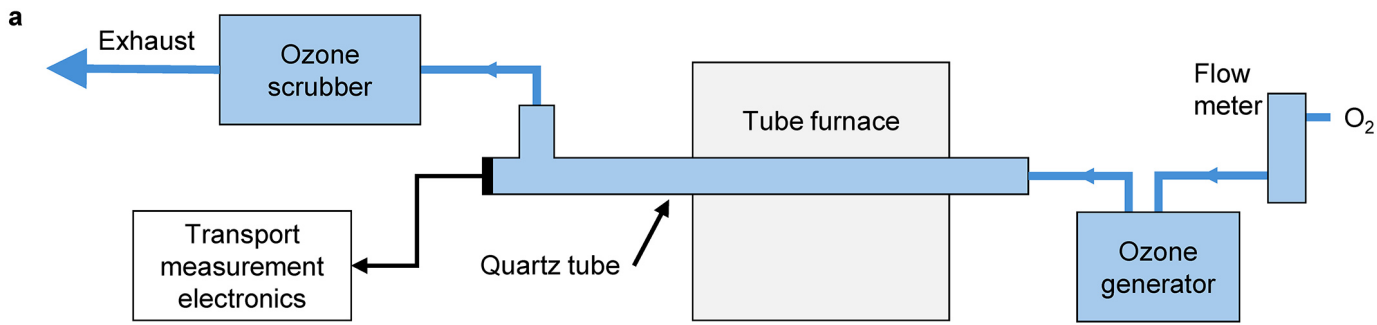

Exhaust — Ozone scrubber — Tube furnace — Flow meter — O₂

Transport measurement electronics — Quartz tube — Ozone generator

**b**

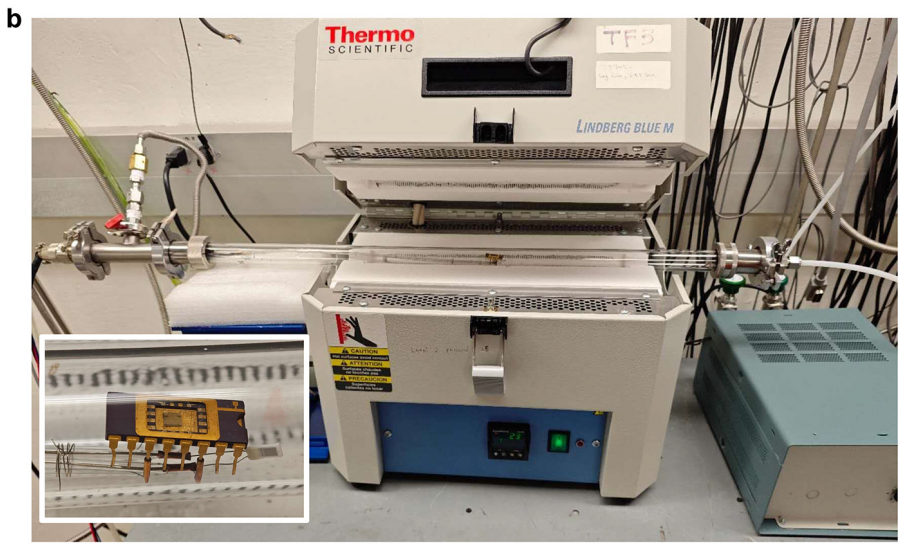

**c**

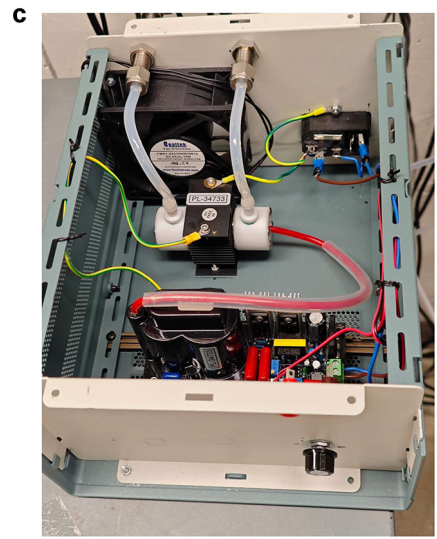

Liu et al., Extended Data Figure 4

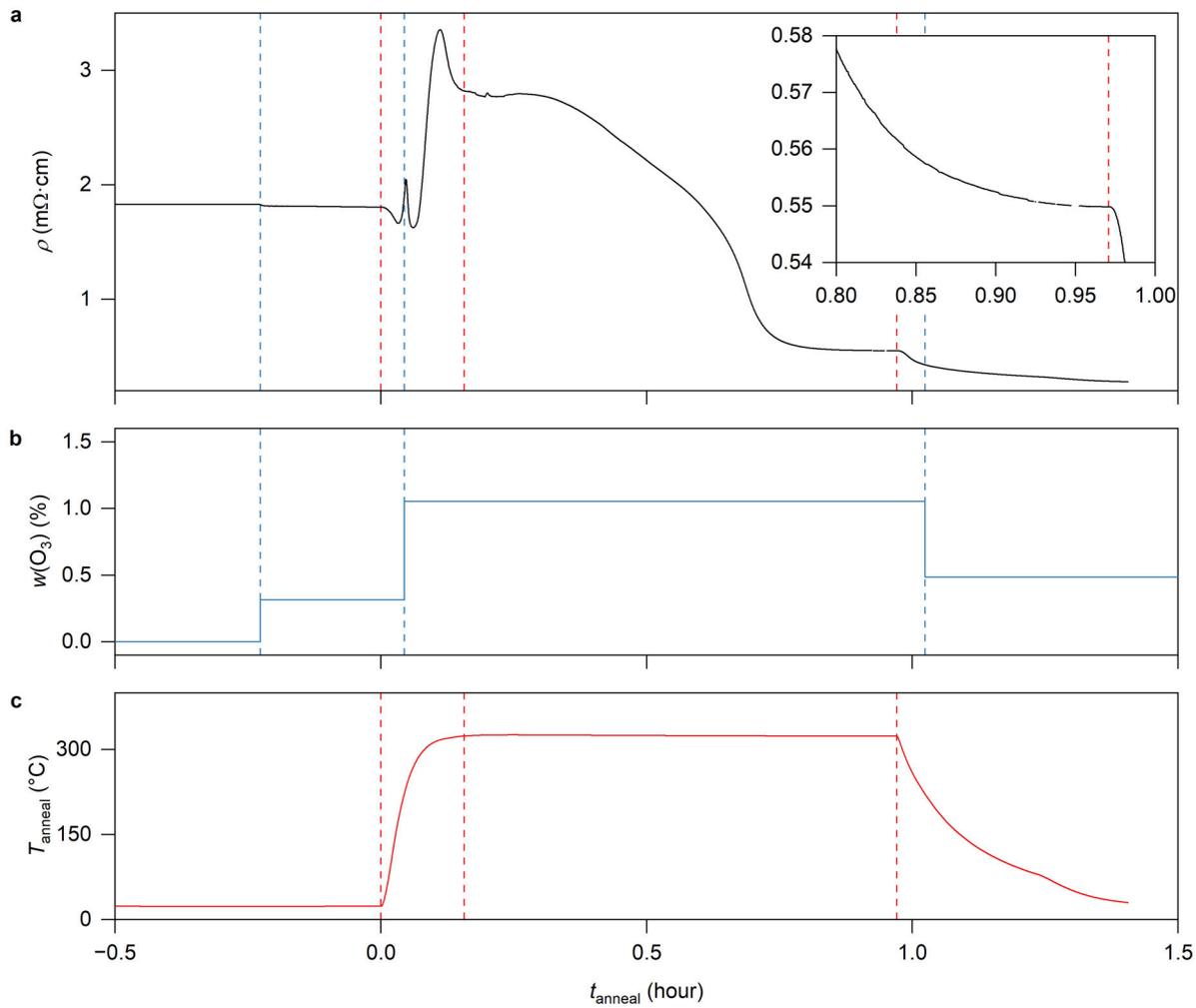

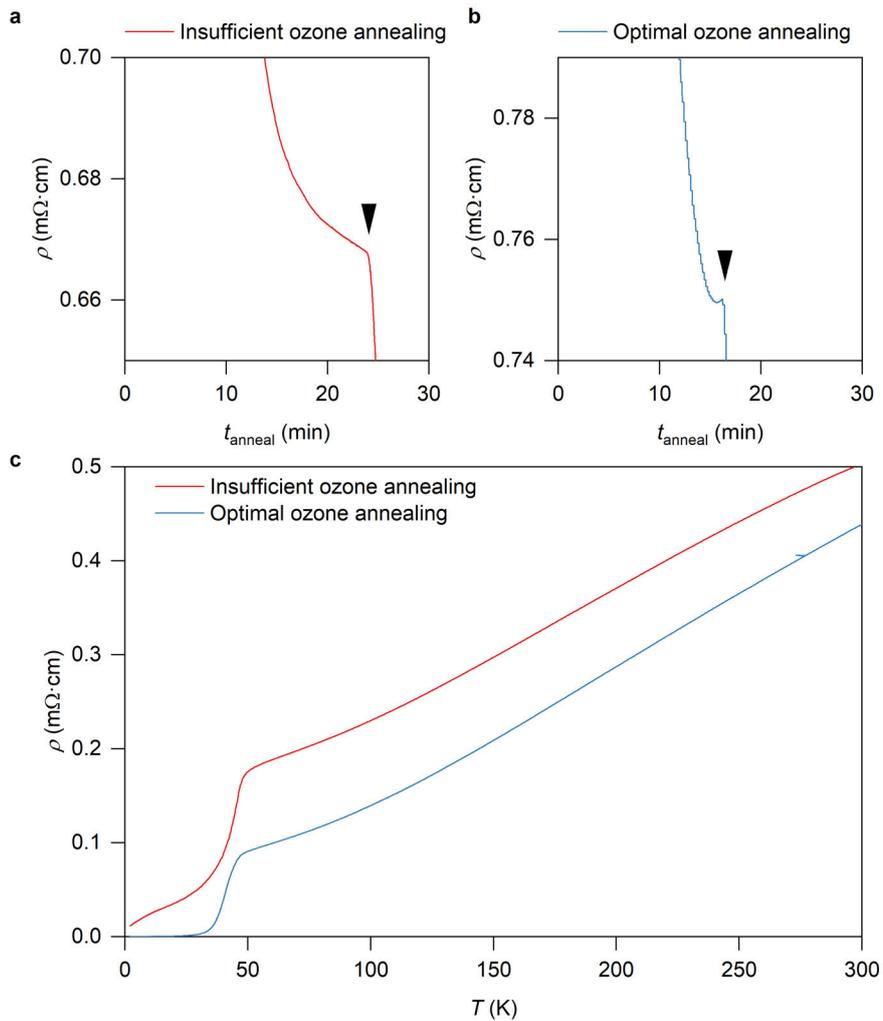

Liu et al., Extended Data Figure 6

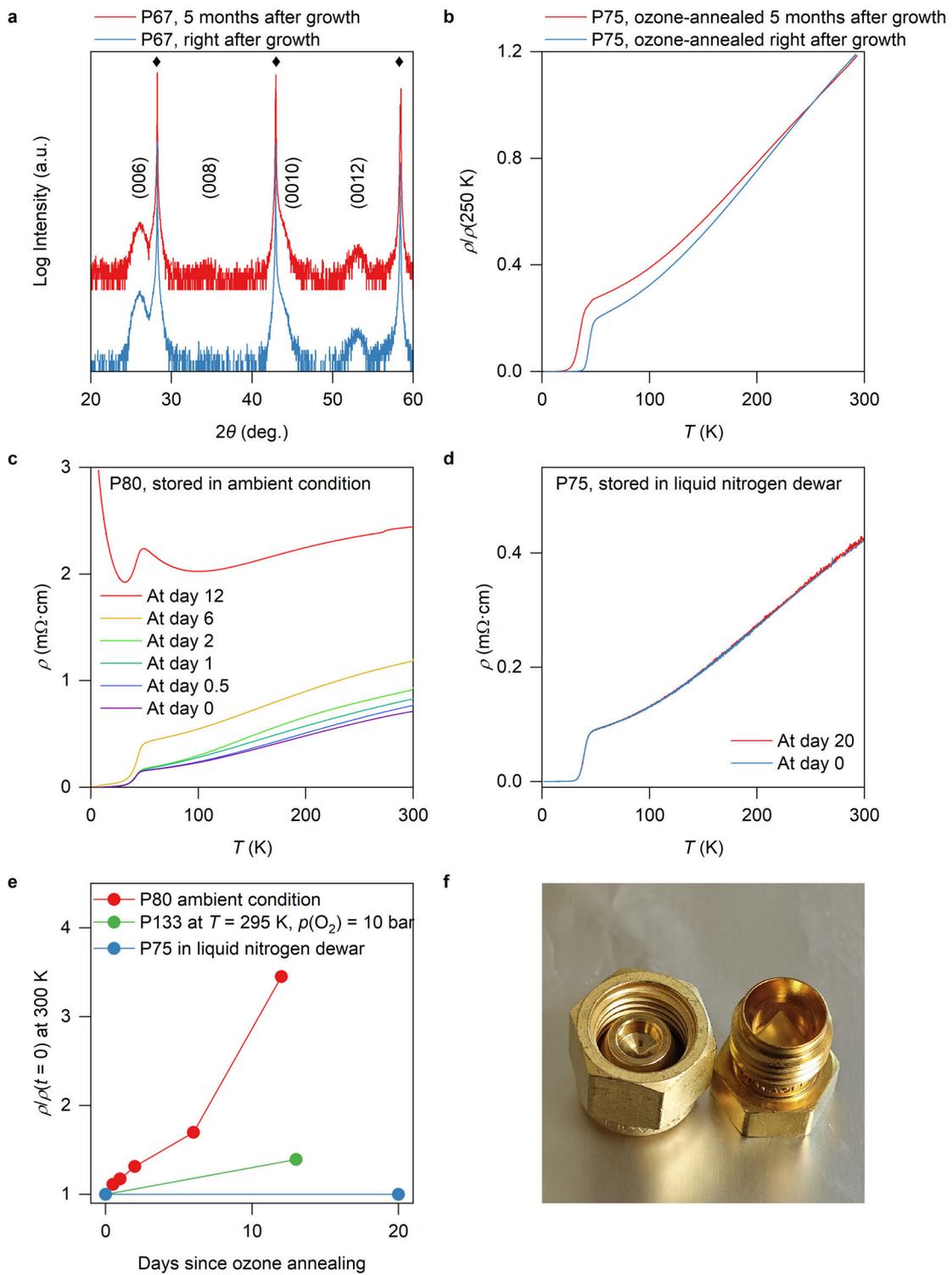

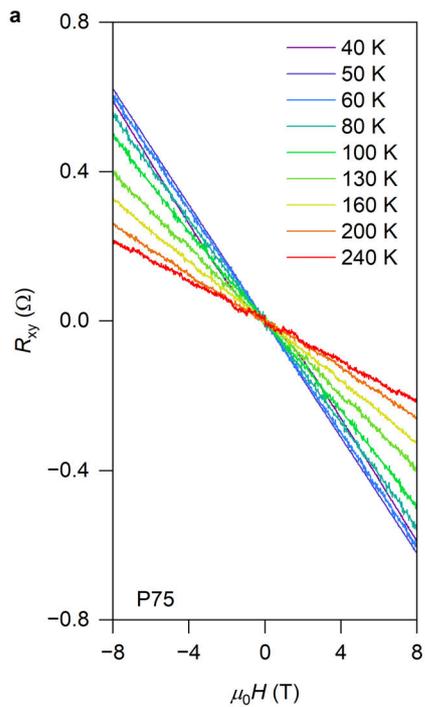
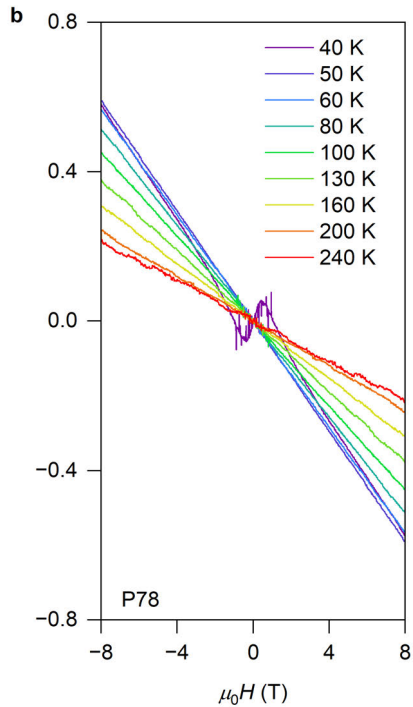
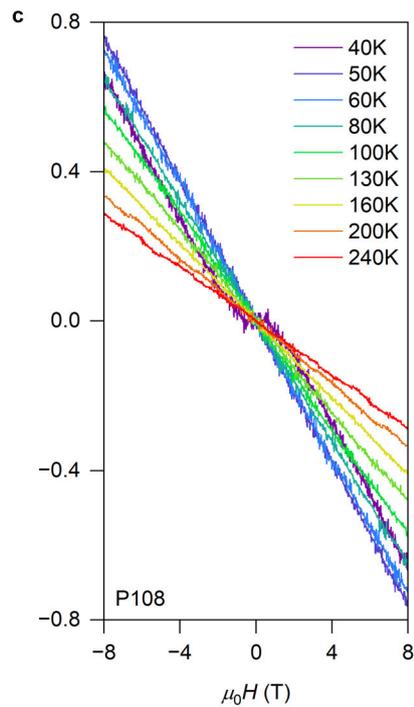

Liu et al., Extended Data Figure 8